\documentclass{article}
\usepackage{amsmath,amssymb, fullpage}

\newcounter {ctr1}
\newtheorem{theorem}[ctr1]{Theorem}
\newtheorem{lemma}[ctr1]{Lemma}

\def\01{\{0,1\}}
\renewcommand{\mathbf}[1]{{\pmb #1}}
\newcommand{\bR}{\mathbb{R}}
\newcommand{\bE}{\mathbb{E}}
\newcommand{\Exp}{\bE}
\newcommand{\Var}{\mbox{\rm Var}}
\newcommand{\eps}{\varepsilon}

\newcommand{\ket}[1]{|#1\rangle}

\DeclareMathOperator{\med}{med}

\begin{document}

\title{Uniform Approximation by (Quantum) Polynomials}
\author{Andrew Drucker%
\thanks{MIT, adrucker@mit.edu.  Supported by a DARPA YFA grant.  The author was supported during part of this work by an Akamai Presidential Graduate Fellowship.} 
\and 
Ronald de Wolf%
\thanks{CWI Amsterdam, rdewolf@cwi.nl.  Partially supported by a Vidi grant from the Netherlands Organization for Scientific Research (NWO), and by the European Commission under the projects Qubit Applications (QAP, funded by the IST directorate as Contract Number 015848) and Quantum Computer Science (QCS).
}}

\date{}
\maketitle

\begin{abstract}
We show that quantum algorithms can be used to re-prove a classical theorem in approximation theory, Jackson's Theorem, which gives a nearly-optimal quantitative version of Weierstrass's Theorem on uniform approximation of continuous functions by polynomials.  We provide two proofs, based respectively on quantum counting and on quantum phase estimation.
\end{abstract}

\section{Introduction}

In many mathematical contexts it is convenient to approximate complicated objects by simpler ones.
A typical example is the approximation of arbitrary continuous functions on a closed interval by polynomials~\cite{Riv}.
Weierstrass's Theorem~\cite{weierstrass} states that this can always be done.
More precisely, the real polynomials are ``dense'' in the space ${\cal C}[0,1]$ of continuous real functions on the unit interval: 
for every $g \in {\cal C}[0,1]$ and $\eps > 0$, there exists a polynomial $p(x)$ such that $|p(x) - g(x)| \leq \eps$ for all $x \in [0,1]$.  

Bernstein~\cite{bernstein:weierstrass} gave a simple and elegant probabilistic construction of such approximating polynomials, which can be described as follows (see also \cite[after Chapter~7]{alon&spencer:probmethod}).
Given a function $g \in {\cal C}[0,1]$ that we want to approximate, 
fix an $n \geq 1$ and flip $n$ independent coins, each coming up `1' with probability $x$.  
Count the Hamming weight $|w|$ of the resulting string $w\in\01^n$, and output $g(|w|/n)$.  
Note that the expected value of $|w|/n$ is exactly $x$, and with high probability we'll have $|w|/n=x\pm O(1/\sqrt{n})$.
But then the output $g(|w|/n)$ should usually be a good estimate of $g(x)$.
Indeed, consider the \emph{expected value} of the output of this algorithm, as a function of $x$:
$$
B_{g,n}(x):=\Exp_w[g(|w|/n)]=\sum_{k=0}^n\frac{{n\choose k}}{2^n} x^k(1-x)^{n-k} g(k/n).
$$ 
This $B_{g,n}$ is a polynomial in $x$ of degree $n$. Since $|w|/n$ is probably close to $x$, we intuitively 
expect $B_{g,n}(x)$ to be close to $g(x)$, provided $g$ does not fluctuate too much on intervals of width $1/\sqrt{n}$.
To capture this fluctuation, define the \textit{modulus of continuity of $g$ at scale} $\delta$ as
$$
\omega_{\delta}(g):=\sup_{x,y:|x-y|\leq\delta}|g(x) - g(y)|.
$$ 
This is a measure of the ``smoothness'' of $g$: the lower the value $\omega_{\delta}(g)$, the more ``smooth'' $g$ is and the smaller our approximation error should be. 
Now an easy argument shows that for every $x\in[0,1]$ we have
$$
|B_{g,n}(x) - g(x)| = O(\omega_{1/\sqrt{n}}(g)),
$$ 
confirming the above intuition.

It is possible to reduce the error of approximation much further.  An improvement of Bernstein's result was shown by Jackson~\cite{jackson:thesis}.  Using trigonometric ideas, he proved

\begin{theorem}[Jackson]\label{thjackson}
There exists a universal constant $C$, such that for every $g\in {\cal C}[0,1]$ and positive integer $n$, 
there is a degree-$n$ polynomial $p$ satisfying $|p(x)-g(x)|\leq C\omega_{1/n}(g)$ for all $x\in[0,1]$.
\end{theorem}

This quality of approximation is based on the maximum fluctuation of $g$ at a much smaller scale than Bernstein's 
($1/n$ instead of $1/\sqrt{n}$).  Up to the constant factor, Jackson's Theorem is optimal for approximation guarantees based on the modulus of continuity. Several different proofs of the theorem are known, see for instance~\cite{bojanovic&devore:jackson, constrapprox}.

In this paper we show how one can implement Bernstein's idea with a \emph{quantum} algorithm, 
improving its error bound to the one in Jackson's Theorem.
Our idea is quite simple: we replace Bernstein's ``algorithm'', which basically counts the number of ones in 
a bitstring of $n$ coin flips, with a quantum counting algorithm.  It was shown by Brassard et al.~\cite{bhmt:countingj} that quantum algorithms can perform (approximate) counting 
more efficiently than classical algorithms, and this yields an improvement over Bernstein's approach.

Let us sketch our proof strategy in somewhat greater detail.  To begin, we perform $N=n^2$ $x$-biased coin flips instead of $n$, yielding a string $w\in\01^N$.  
Now we have $|w|/N=x\pm O(1/\sqrt{N})=x\pm O(1/n)$ with high probability, 
so $|w|/N$ is a much more precise estimator of $x$ than the $|w|/n$ of Bernstein's proof.
We then run a quantum counting algorithm making $n/2$ ``quantum queries'' to $w$.  This algorithm computes an estimator $A$ of $|w|/N$, such that with high probability
$$
|A - |w|/N|\leq O(1/n).
$$
Accordingly, with high probability $A$ also approximates $x$ within error $O(1/n)$.
Then intuitively the function 
$$
Q_{g,N}(x):=\Exp_{w,A}[g(A)]
$$
should approximate $g$ up to error roughly $\omega_{1/n}(g)$.  The expectation in the above expression is over both the choice of $w$,
and over the randomness generated by the measurements in the quantum counting algorithm that runs on $w$ and outputs $A$.

It is well known that the acceptance probability of an $n/2$-query quantum algorithm is an $N$-variate multilinear
polynomial (in the bits of $w$) of degree at most $n$ (see~\cite{bbcmw:polynomialsj}). Taking the expectation over $w$ turns this into
a univariate polynomial in $x$ of degree at most $n$, since the expectation of a monomial $w_{i_1}\cdots w_{i_d}$ is exactly $x^d$.
Working out the details (Section~\ref{secjacksoncount}), $Q_{g,N}$ is indeed a degree-$n$ polynomial
that approximates $g$ within error bound $O(\omega_{1/n}(g))$.  This reproves Jackson's theorem.

As we explain in the next section, efficient quantum counting relies on a more basic procedure called \emph{quantum phase estimation}~\cite{kitaev:stabilizer}.
In Section~\ref{secjacksonphase} we give a second proof of Jackson's Theorem that is directly based on phase estimation.
This naturally yields a proof of the \emph{trigonometric} version of Jackson's Theorem  (Theorem~\ref{thjacksontrig} below), but in fact the two versions are equivalent (see~\cite{Riv}).
A \textit{trigonometric polynomial} of degree $n$ can be defined in two equivalent ways: 
as a sum of the form $p(x)=\sum_{k = -n}^{n} \alpha_k e^{2\pi i x k}$, or as a sum of the form 
$p(x)=a_0 + \sum_{k = 1}^n \left(a_k \cos (2\pi k x)  +   b_k \sin (2\pi k x) \right)$, where $\alpha_k, a_k, b_k$ are scalars, possibly complex.%
\footnote{The equivalence of the definitions follows from Euler's formula $e^{i\theta}  = \cos \theta + i \sin \theta$.}  
A trigonometric polynomial is \textit{real} if it maps real numbers to real numbers, or equivalently, if the coefficients $a_k, b_k$ are all real.
Note that a trigonometric polynomial is a 1-periodic continuous function.  
The trigonometric polynomials are dense in the set of all such functions, and the trigonometric version of Jackson's Theorem gives a quantitative refinement of this fact:

\begin{theorem}[Jackson, trigonometric version]\label{thjacksontrig}
There exists a universal constant $C$, such that for every 1-periodic function $g\in {\cal C}[\bR]$ and positive integer $n$, 
there is a degree-$n$ real trigonometric polynomial $p$ satisfying $|p(x)-g(x)|\leq C\omega_{1/n}(g)$ for all $x\in\bR$.
\end{theorem}

The best constant was determined by Korneichuk, who showed that every 1-periodic $g\in {\cal C}[\bR]$ 
can be approximated by a degree-$n$ trigonometric polynomial with error $\eps<\omega_{1/2n}(h)$, which is essentially optimal.%
\footnote{By~\cite[Theorem~6.2.2]{korneichuk:exact}, every $2\pi$-periodic $h\in {\cal C}[\bR]$ can be approximated
by a degree-$n$ trigonometric polynomial with error $\eps<\omega_{\pi/n}(h)$; we have restated this here for 1-periodic functions.
The error bound is essentially optimal: for every $n$ and $\alpha>0$, there is a $2\pi$-periodic function $h$ such that every degree-$n$ polynomial
differs from $h$ by at least $(1-1/2n-\alpha)\omega_{\pi/n}(h)$ (see~\cite[Lemma~6.2.3]{korneichuk:exact}).}

Though the ideas used in our two approaches to Jackson's Theorem are closely related,
we feel both have merit. The one based on quantum counting (Section~\ref{secjacksoncount}) is a quantum generalization
of Bernstein's proof, while the one based on phase estimation (Section~\ref{secjacksonphase}) 
is a more ``direct'' approach since phase estimation is the basis for quantum counting.
The phase estimation approach is in fact closely related to Jackson's original proof, as we explain in Section~\ref{secrelationcla}.  

Finally, let us mention that this paper fits in a sequence of recent applications of quantum computational techniques 
establishing or casting new light on results that have nothing to do with quantum computing itself.
We refer to~\cite{drucker&wolf:qproofs} for a survey.

\section{Preliminaries: Phase estimation and quantum counting}\label{app:appA}

Here we sketch how phase estimation works and how it can be used to do approximate quantum counting.
The presentation is based on Brassard et al.~\cite{bhmt:countingj}; the phase estimation algorithm is due to Kitaev~\cite{kitaev:stabilizer}.

\subsection{Phase estimation}\label{subsec:phasesec}
Suppose we can apply a certain unitary $U$ as often as we want, and we are given one of $U$'s eigenvectors,
$\ket{u}$, with unknown eigenvalue $e^{2\pi i x}$ for~$x\in[0,1)$. We would like to learn~$x$.
Quantum phase estimation allows us to approximate $x$ up to any desired precision, as described below.
It will be convenient to define the following distance for approximations: 
$d(\tilde{x},x):=\min_{c\in\mathbb{Z}}|c+x-\tilde{x}|\in[0,1/2]$, so $2\pi d(\tilde{x},x)$ 
is the shortest distance along the unit circle from $e^{2\pi i x}$ to $e^{2\pi i \tilde{x}}$.

We start with a 2-register quantum state, where the first register contains the uniform superposition over $M$ basis states, 
and the second contains the eigenvector $\ket{u}$:
$$
\frac{1}{\sqrt{M}}\sum_{y=0}^{M-1}\ket{y}\otimes\ket{u}.
$$
Now, conditioned on the first register's value $y$, apply $U$ to the second register $y$ times, 
i.e., map $\ket{y}\otimes\ket{u}\mapsto e^{2\pi i x y}\ket{y}\otimes\ket{u}$.  This gives
$$
\frac{1}{\sqrt{M}}\sum_{y=0}^{M-1}e^{2\pi i x y}\ket{y}\otimes\ket{u}.
$$
From now on we ignore the second register.
Apply the inverse quantum Fourier transform over $\mathbb{Z}_M$ to get
$$
\frac{1}{\sqrt{M}}\sum_{y=0}^{M-1}e^{2\pi i x y}\frac{1}{\sqrt{M}}\sum_{z=0}^{M-1}e^{-2\pi i yz/M}\ket{z}
=\sum_{z=0}^{M-1}\alpha_z\ket{z},\mbox{ where }\alpha_z=\frac{1}{M}\sum_{y=0}^{M-1} e^{2\pi i y d(z/M,x)}.
$$
If we measure this, we get a random variable $Z\in\{0,\ldots,M-1\}$ with distribution
\begin{equation}\label{eqerrorphasest}
\Pr[Z=z]=|\alpha_z|^2=
%\frac{|1-e^{2\pi i M d(z/M,x)}|}{M^2|1-e^{2\pi i d(z/M,x)}|}
\left\{\begin{array}{ll} 1 & \mbox{if $d(z/M,x)=0$,}\\
                           \frac{\sin(M d(z/M,x)\pi)^2}{M^2\sin(d(z/M,x)\pi)^2} & \mbox{otherwise,}\end{array}\right.
\end{equation}
where we used the identities $\sum_{y=0}^{M-1}r^y=(1-r^M)/(1-r)$ and $|1-e^{i\phi}|=2|\sin(\phi/2)|$.
Note that the first case is just the limit of the second case as $d(z/M,x)\rightarrow 0$.
The distribution $\Pr[Z=z]$ is peaked at values of $z$ where $d(z/M,x)$ is small, so we can use $\tilde{X}:=Z/M$ as our estimate of $x$.
The probability of outcome $\tilde{X} = \tilde{x}$ falls off quadratically with its distance from $x$: 
if $\tilde{x} \neq x$ then  (using $\sin(\phi)\geq 2\phi/\pi$ for $\phi\in[0,\pi/2]$)
\begin{equation}\label{eqphasesterror}
\Pr[\tilde{X}=\tilde{x}]=\Pr[Z=\tilde{x}M]\leq\frac{1}{4M^2 d(\tilde{x},x)^2}.
\end{equation}
For convenience, we let $1/0 = \infty$ and consider Eq.~(\ref{eqphasesterror}) to hold vacuously when $d(\tilde{x}, x)= 0$.

\subsection{Quantum counting}
Suppose we have an input $w\in\01^N$ whose Hamming weight $|w|$ we want to estimate, using a quantum query algorithm that can access $w$ by means of the query operator $O_w$, which maps $\ket{i}\mapsto(-1)^{w_i}\ket{i}$.
Then we can perform approximate counting using phase estimation as follows.

For simplicity assume $N=2^n$.
Define a unitary $U=-H^{\otimes n}O_0H^{\otimes n}O_w$, 
where $O_0$ is the unitary that puts a `$-$' in front of the all-0 state, and $H$ is the Hadamard transform.
This $U$ is known as the \emph{Grover iterate}, and is the crucial ingredient in the quantum search algorithm~\cite{grover:search,bhmt:countingj}.
Let $\ket{\Psi_1}=\frac{1}{\sqrt{|w|}}\sum_{i:w_i=1}\ket{i}$ and $\ket{\Psi_0}=\frac{1}{\sqrt{N-|w|}}\sum_{i:w_i=0}\ket{i}$
be the uniform superpositions over the 1-bits and the 0-bits of $w$, respectively.
One can show~\cite{bhmt:countingj} that $U$ has the following two orthogonal eigenvectors, with corresponding eigenvalues:
\begin{quote}
$\displaystyle\ket{\Psi_+}=\frac{1}{\sqrt{2}}\left(\ket{\Psi_1}+i\ket{\Psi_0}\right)$ with $\lambda_+=e^{2i\theta}$,\\
$\displaystyle\ket{\Psi_-}=\frac{1}{\sqrt{2}}\left(\ket{\Psi_1}-i\ket{\Psi_0}\right)$ with $\lambda_-=e^{-2i\theta}$,\\
where $\theta=\arcsin(\sqrt{|w|/N})\in[0,\pi/2]$.
\end{quote}
Note that $|w|/N=\sin(\theta)^2$. We want to estimate $\theta$ by means of phase estimation on $U$.
For this we would need an eigenvector of $U$ with eigenvalue related to $\theta$.
We cannot easily construct one of the two eigenvectors $\ket{\Psi_+}$ and $\ket{\Psi_-}$.
However, the uniform superposition 
$$
\ket{u}=\frac{1}{\sqrt{N}}\sum_{i=1}^N\ket{i}=\sqrt{\frac{|w|}{N}}\ket{\Psi_1}+\sqrt{\frac{N-|w|}{N}}\ket{\Psi_0}
$$
is a linear combination of $\ket{\Psi_+}$ and $\ket{\Psi_-}$.
This $\ket{u}$ is independent of $w$ and easy to construct.
We can analyze this as if the second register contains a \emph{mixture} of the two eigenvectors. Thus, doing phase estimation on $U$ with starting vector $\ket{u}$, 
we will be estimating either $\theta$ or $-\theta$.  Since $\sin(\theta)^2=\sin(-\theta)^2$, 
we don't care whether we estimate $\theta$ or $-\theta$ (assume the first one for simplicity).
Phase estimation with $\ket{u}$ as starting vector produces a random variable $Z\in\{0,\ldots,M-1\}$, 
such that the distribution of $Z/M$ is peaked around $\theta/\pi$.
Accordingly, we use $\tilde{\theta}:=\pi Z/M$ as our estimate of $\theta$, and $A :=\sin(\tilde{\theta})^2$ as our estimator of $|w|/N=\sin(\theta)^2$.
Note that the number of queries of this procedure is the number of applications of $U$, which is $M-1$.
Eq.~(\ref{eqphasesterror}) gives a tradeoff between the number of queries and the error in our approximation of $\theta$, 
which translates into an error in our approximation of $|w|/N$.
Brassard et al.~\cite{bhmt:countingj} work out various points on this tradeoff in detail.

\section{Jackson's Theorem by quantum counting}\label{secjacksoncount}

In this section we provide the details of the idea sketched in the Introduction, 
combining Bernstein's probabilistic approach with a quantum counting algorithm.  We first present a direct approach that does not quite work; we then explain how a simple modification allows the proof to go through.

We want to construct an approximating polynomial of degree $n$ for a given continuous function $g\in {\cal C}[0,1]$.
We start by setting $N :=n^2$ and letting $w\in\01^N$ be a string obtained by flipping $N$ coins, each with probability $x$ of `1'.
Since its Hamming weight $|w|$ is binomially distributed with expectation $xN$ and variance $x(1-x)N$, we have 
$\Exp[| x-|w|/N|]\leq \sqrt{\Var[|w|/N]}=\sqrt{x(1-x)/N}=O(1/n)$.
We now want to estimate $|w|/N$ using quantum counting.
As explained in Section~\ref{app:appA}, we define $\theta:=\arcsin(\sqrt{|w|/N})$, so that $\sin(\theta)^2=|w|/N$.
Letting $M > 1$ be an integer to be fixed later, quantum counting with $M-1$ queries produces a random variable 
$Z\in\{0,\ldots,M-1\}$, such that by Eq.~(\ref{eqphasesterror}),
$$
\Pr[Z=z]\leq\frac{1}{4M^2d(z/M,\theta/\pi)^2},
$$
where $d(\cdot,\cdot)$ is as defined in Section~\ref{subsec:phasesec}.
Since $Z/M$ is concentrated (with respect to the distance $d(\cdot,\cdot)$)
around $\theta/\pi$, we can use $\tilde{\theta}:=\pi Z/M$ as our estimate of $\theta$, 
and $A:=\sin(\tilde{\theta})^2$ as our estimate of $|w|/N$. 
We claim that the angles $\tilde{\theta}, \theta$ are likely to be close ``modulo $\pi$'': we have
\begin{align}\label{eqcountest}
 \displaystyle   \Exp[ d(\tilde{\theta}/\pi , \theta/\pi)] =\sum_{z=0}^{M-1}\Pr[Z=z]\cdot d(z/M,\theta/\pi) &\leq \frac{2}{M} + \sum_{\substack{z\in\{0,\ldots,M - 1\},\\|z - M\theta/\pi| \geq 1} }\frac{1}{4M^2d(z/M,\theta/\pi)} \nonumber \\
&=O\left(\frac{1}{M} +\frac{1}{M}\sum_{t=1}^M\frac{1}{t}\right)  =O\left(\frac{\log M}{M}\right).
\end{align}
The function $f(t) = \sin (t)^2$ is $\pi$-periodic and its derivative satisfies $|f'(t)| = |\sin(2t)| \leq 1$.  Thus
\begin{equation}\label{eqsine}
|\sin(\tilde{\theta})^2-\sin(\theta)^2|\leq \pi\cdot d(\tilde{\theta}/\pi,\theta/\pi),
\end{equation}
and hence also 
\begin{equation}\label{eqAerr}
\Exp[|A-|w|/N|]=\Exp[|\sin(\tilde{\theta})^2-\sin(\theta)^2|] \leq \pi \cdot \Exp[ d(\tilde{\theta}/\pi,\theta/\pi)]  = O((\log M)/M).
\end{equation}
Define a function $p: [0, 1] \rightarrow \bR$ by
\begin{equation}\label{eqdefp}
p(x):=\Exp_{w,A}[g(A)]=\sum_a\Exp_w[\Pr[A=a\mid w]] \cdot g(a),
\end{equation}
where $\Pr[w]=x^{|w|}(1-x)^{N-|w|}$ depends on $x$, and $\Pr[A=a\mid w]$ is the probability that quantum counting on input $w$ yields estimate $a$ for $|w|/N$.
The following lemma from~\cite{bbcmw:polynomialsj} implies that for any fixed value $a$, $\Pr[A=a\mid w]$ is a low-degree polynomial in the $N$ variables of $w$:

\begin{lemma}[BBCMW]\label{lem:degree}
Consider a quantum algorithm that makes at most $T$ queries to $w\in\01^N$ and outputs the result of a measurement on the final state.
Then the probability $\Pr[A=a\mid w]$ of any specific output $a$ is an $N$-variate multilinear real polynomial in $w$ of degree at most $2T$.
\end{lemma}

Each polynomial $\Pr[A=a\mid w]$ is a linear combination of monomials in $w_1,\ldots,w_N$, 
of degree at most $2T$.
A degree-$d$ monomial $w_{i_1}\cdots w_{i_d}$ has expectation $\Exp_w[w_{i_1}\cdots w_{i_d}]=\prod_{j=1}^d\Exp_w[w_{i_j}]=x^d$, 
which is a polynomial in $x$ of degree $d$.
Hence each expression $\Exp_w[\Pr[A=a\mid w]]$ in Eq.~(\ref{eqdefp}) is a polynomial in $x$ of degree at most $2M-2$.  
Then the function $p$ defined in Eq.~(\ref{eqdefp}) 
is itself also a polynomial in $x$ of degree at most $2M-2$.  Choosing $M:=n/2+1$, $p$ has degree at most $n$.

Next we argue that $p$ approximates $g$ fairly well.  First, recall the definition of $\omega_{\delta}(g)$ from the Introduction; we have the elementary property $\omega_{\delta+\delta'}(g)\leq\omega_{\delta}(g)+\omega_{\delta'}(g)$, and therefore $\omega_{c \delta}(g) \leq \lceil c \rceil \cdot \omega_{\delta}(g)$.
Now, for every $x\in[0,1]$ we bound
\begin{eqnarray}\label{eqineqs}
|p(x)-g(x)| & \leq & \Exp_{w,A}[ |g(A)-g(x)| ] \nonumber\\
            & \leq & \Exp_{w,A}[ \omega_{|A-x|}(g) ]\nonumber\\
            & \leq & \Exp_{w,A}[ \omega_{|A-|w|/N|}(g) + \omega_{|x-|w|/N|}(g) ] \nonumber\\
            & \leq &  \Exp_{w,A}\left[ \left\lceil \left|A-|w|/N \right| \cdot (n / \log n ) \right\rceil \right]\omega_{\log n/ n}(g) + \Exp_{w}\left[ \left\lceil \left|x-|w|/N\right| \cdot n \right\rceil \right]\omega_{1/n}(g)\nonumber\\
            & \label{niceineq}\leq & O(\omega_{\log n /n}(g) + \omega_{1/n}(g))\nonumber\\
            & \leq & O(\omega_{\log n /n}(g)),
\end{eqnarray}
where in the penultimate step we used that $\bE[|A-|w|/N |] = O((\log n)/n)$ and $\bE[|x - |w|/N|] = O(1/n)$.
The error bound we derived is already substantially better than Bernstein's bound of $O(\omega_{1/\sqrt{n}}(g))$, but worse than Jackson's optimal error 
bound $O(\omega_{1/n}(g))$ by a logarithmic factor. To get rid of this factor, we want an estimate of $|w|/N$ with expected error $O(1/n)$ instead of $O((\log n)/n)$.

Taking an algorithmic perspective, it is easy to see how to get a more sharply concentrated estimate.  We run the quantum counting procedure three times, yielding estimates $Z_1,Z_2,Z_3$ for $\theta M/\pi$.  These yield estimates $\tilde{\theta}_i := \pi Z_i/M$ for $\theta$, for $i = 1, 2, 3$.  We then get estimates $A_1, A_2, A_3$ for $|w|/N$, by taking $A_i := \sin(\tilde{\theta}_i )^2$.  We let $A' := \med(A_1, A_2, A_3)$ be defined as the \emph{median} of these three estimates, and use $A'$ as our improved estimate for $|w|/N$.  In order to keep the total number of queries at most $n/2$ (and hence the degree at most $n$), we will now spend $M:=n/6$ queries for each of the three runs of quantum counting, assuming for simplicity that 6 divides $n$.

We now show that this approach yields a tighter estimate.  First, it can be easily verified that for real variables $a, b, c, t$,  
\begin{equation}\label{eqbasicmed}
|\med(a, b, c) - t| \leq \med(| a - t|, | b - t|, | c - t|).
\end{equation}
It follows
that $|A' - |w|/N|$ is at most the median of $|A_i - |w|/N|$, over $i = 1, 2, 3$.  Using Eq.~(\ref{eqsine}) applied to each of the $\tilde{\theta}_i$, we obtain $|A_i - |w|/N| \leq \pi \cdot d(\tilde{\theta}_i/\pi, \theta/\pi)$.
If we define a random variable $d_{med}$ as the median of the quantities $d(\tilde{\theta}_i/\pi, \theta/\pi)$, for $i = 1, 2, 3$, then we have
\begin{equation}\label{eqmed-bound}
|A' - |w|/N|  \leq \pi \cdot d_{med}.
\end{equation}
We now bound $\bE[d_{med}]$.  Fix an integer $k \geq 1$.  For $d_{med} \in [k/M, (k+1)/M)$ to occur, it is necessary that some $Z_i$ satisfies $d(Z_i/M, \theta/\pi) \in [k/M, (k+1)/M)$, and some $Z_j$ with $j \neq i$ satisfies $d(Z_j/M, \theta/\pi) \geq k/M$.  There are at most two possible values $z \in \{0, 1, \ldots, M-1\}$ satisfying $d(z/M, \theta/\pi) \in [k/M, (k+1)/M)$, and there are just 6 possible pairs $i, j$.  Thus, we have
\begin{align}
\Pr \left[d_{med} \in [k/M, (k+1)/M) \right] &\leq \nonumber\\
 6 \cdot \Pr \left[d(Z_1/M, \theta/\pi) \in [k/M, (k+1)/M) \right]  &\cdot \sum_{\substack{z':\\  d(z'/M, \theta/\pi) \geq k/M}} \Pr[Z_2 = z'] \nonumber \\
&\leq O \left(  \frac{1}{  M^2(k/M)^2}  \sum_{k' \geq k} \frac{1}{  M^2(k'/M)^2}    \right) \quad{} \text{(using Eq.~(\ref{eqphasesterror}))} \nonumber \\
&\leq  O\left( \frac{1}{k^2} \sum_{k' \geq k}  \frac{1}{(k')^2} \right)\nonumber\\
&\leq O\left( \frac{1}{k^3}  \right).\nonumber
\end{align}%  \Pr[d(Z_2/M, \theta/\pi) \geq k/M]
It follows that 
$$
\bE[d_{med}] \leq \frac{1}{M} + \sum_{k \geq 1}\frac{k+1}{M}\cdot O\left( \frac{1}{k^3}  \right)  =  \frac{1}{M}\cdot O\left(1 + \sum_{k \geq 1} \frac{1}{k^2}  \right) = O\left(1/M \right) = O(1/n),
$$
and by Eq.~(\ref{eqmed-bound}) we get $\bE[|A' - |w|/N|] = O\left(1/n\right)$.

We redefine the polynomial $p$ as $p(x) := \bE_{w, A'}[g(A')]$.  By our revised setting $M = n/6$, $p(x)$ is a polynomial of degree $\leq n$.  Following the steps of Eq.~(\ref{niceineq}) with appropriate changes, we have 
\begin{align}
|p(x)-g(x)| &\leq \Exp_{w,A'}[ \omega_{|A'-|w|/N|}(g) + \omega_{|x-|w|/N|}(g) ] \nonumber\\
            &\leq   \Exp_{w,A'}\left[ \left\lceil |A'-|w|/N | \cdot n \right\rceil \right]\omega_{1/n}(g) + O(\omega_{1/n}(g))  \nonumber\\
           &\leq  O(\omega_{1 /n}(g)).\nonumber
\end{align}
This proves Theorem~\ref{thjackson}.

\section{Jackson's Theorem by phase estimation}\label{secjacksonphase}

As explained in Section~\ref{app:appA}, quantum counting is based on quantum phase estimation.
In this section we describe an alternative way to prove (the trigonometric version of) Jackson's theorem.
We will no longer estimate the weight of a string of $x$-biased coin flips, but instead apply quantum phase estimation directly to a unitary which ``encodes'' $x$.

Suppose we apply phase estimation to the $1\times 1$ unitary $U=[e^{2\pi i x}]$ to estimate $x$.
Since $U$ is 1-dimensional, every vector $\ket{u}$ is an eigenvector with eigenvalue $e^{2\pi i x}$.
As explained in Section~\ref{app:appA}, using up to $M-1$ applications of $U$,
phase estimation produces a $Z\in\{0,\ldots,M-1\}$, distributed as in Eq.~(\ref{eqerrorphasest}), so that $Z/M$ is concentrated around $x$ (with respect to the distance $d(\cdot, \cdot)$).  Let $M := n/3 + 1$, assuming for simplicity that 3 divides $n$.  Suppose we apply phase estimation three times, getting outcomes $Z_1,Z_2, Z_3\in\{0,\ldots,M-1\}$.
Define $Y$ as the median of $g(Z_i/M)$ over $i = 1, 2, 3$; we use $Y$ as an estimate of $g(x)$.  Note how we now take medians \emph{after} applying $g$, in contrast to our previous approach.\footnote{We could have taken medians after applying $g$ in Section~\ref{secjacksoncount} as well, but we wanted to keep a conceptual focus on obtaining a sharp estimate of the quantity $|w|/N$ (in analogy with Bernstein's proof).  In the phase estimation approach, sharpening our estimate of $x$ would be slightly more involved, and taking medians after applying $g$ yields a more streamlined proof.}

In order to approximate the continuous 1-periodic function $g\in{\cal C}[\bR]$, consider the quantum algorithm that produces an estimate $Y$ as above, using the unitary $U=[e^{2\pi i x}]$.  We define $p(x) := \bE[Y]$.
Referring back to our description of phase estimation, we see that the amplitudes of the final state of our quantum algorithm are linear combinations of $e^{2\pi i x k}$ for $k\in\{0,\ldots,3(M-1)\}$.
Hence the final measurement probabilities are trigonometric polynomials in $x$ of degree $3(M-1)= n$.\footnote{While the \emph{amplitudes} are linear combinations of $e^{2\pi i x k}$ for $k\in\{0,\ldots,3(M-1)\}$, the \emph{probabilities} are sums of amplitudes times their conjugates, and hence are linear combinations of $e^{2\pi i x k}$ with $k\in\{-3(M-1), \ldots, 3(M-1)\}$.}  
Accordingly, $p$ is such a polynomial as well.

Since the distribution of each $Z_i$ is concentrated around $x$ (with respect to $d(\cdot, \cdot)$) and $g$ is continuous and 1-periodic, we expect $p(x)$ to be close to $g(x)$.  We make this precise next.
For each $i$, the definition of $\omega_{\delta}(g)$ and the fact that $g$ is 1-periodic implies that
\begin{equation}\label{eqomegabd}
|g(Z_i/M) - g(x)| \leq  \left\lceil  d(Z_i/M, x) \cdot  n  \right\rceil   \omega_{1/n}(g) .
\end{equation}
Eq.~(\ref{eqbasicmed}) implies that $|Y - g(x)|$ is at most the median of $|g(Z_i/M) - g(x)|$, over $i = 1, 2, 3$.  By analogy with the previous section, define the random variable $d_{med}$ as the median of $d(Z_i/M, x)$, $i = 1, 2, 3$.  From Eq.~(\ref{eqomegabd}), we obtain
\begin{equation}\label{eqdmed}
|Y - g(x)| \leq \left\lceil  d_{med} \cdot n  \right\rceil    \omega_{1/n}(g).
\end{equation}
Reasoning identical to that of the previous section yields $\bE[d_{med}] = O(1/M) = O(1/n)$, and therefore 
\begin{align*}
|p(x) - g(x)| &\leq \bE \left[ |Y - g(x)|  \right] \\
&\leq \bE\left[   \left\lceil  d_{med} \cdot n  \right\rceil      \right]  \omega_{1/n}(g)\\
&= O(\omega_{1/n}(g)).
\end{align*}
This proves Theorem~\ref{thjacksontrig}, the trigonometric version of Jackson's Theorem.

\section{Relation to classical proofs}\label{secrelationcla}

Our approach to Jackson's Theorem in both sections bears strong similarities with classical proofs.  The first approach, using quantum counting, was modelled on the probabilistic interpretation of Bernstein's polynomial approximation.  The second approach, using phase estimation, also has a very close relation to a classical technique in approximation theory, the method of \textit{convolution with an approximation kernel}~\cite{SS}.  This method was employed by Jackson in his original proof~\cite{jackson:thesis}.  We explain this method and its relation to our proof next.

Suppose $K(t) \in {\cal C}(\bR)$, the ``approximation kernel,'' is a nonnegative, $1$-periodic function concentrated around zero (mod $1$), and such that $\int_{0}^{1}K(t)dt = 1$.  Then for any $x \in \bR$, the function $K_x(t) := K(t - x)$ is concentrated around $x$ (mod $1$).  This suggests that, for a 1-periodic function $h(t) \in {\cal C}(\bR)$, the \textit{convolution} of $h$ with $K$, i.e.,
$$
(h * K)(x) := \int_{0}^{1}h(s)K(s - x)ds,
$$ 
should give a good approximation to $h(x)$.  As for degree considerations, a key fact is that, if $K$ is a degree-$n$ trigonometric polynomial, then so is $(h * K)$ (see, e.g., \cite[Chap.~3]{SS}).
Finally, note that the expression defining $(h * K)(x)$ is exactly $\bE[h(\tilde{x})]$, when $\tilde{x} \in [0, 1)$ is an approximation to $x$ drawn according to the density function $K_x$.  Thus the convolution approach resembles the phase-estimation approach (with a single application of phase estimation), except that an estimator with a continuous distribution is used.\footnote{Alternatively, the phase-estimation approach can also be reinterpreted as using a \emph{discrete convolution}, once this notion is made formal.}    

In fact, even the distribution of the estimates we derive from phase estimation is intimately related to the approximation kernels used in Jackson's original proof \cite{jackson:thesis, lorentz, constrapprox}.  A classical approximation kernel is the \emph{Fej\'{e}r kernel} $F_n: \mathbb{R} \rightarrow \mathbb{R}$ ($n > 0$ is an integer parameter), given by
$$
F_n(t) = \frac{1}{n}\left(\frac{\sin (\pi nt)}{\sin (\pi t)}\right)^2,
$$
with $F_n(0) = n$ for continuity.  Note that $F_n$ as defined here is 1-periodic (it is usually defined in a $2\pi$-periodic form).  Also, $F_n$ can be re-expressed as a trigonometric polynomial of degree $n - 1$. Now we compare this to our analysis of phase estimation in Section~\ref{app:appA}.  Consulting Eq.~(\ref{eqerrorphasest}), if the unknown eigenvalue $e^{2\pi i x}$ ($x \in [0, 1)$) satisfies $Mx \notin \mathbb{Z}$, then the estimate 
$\tilde{y} = Z/M$ to $x$ produced by phase estimation ($z\in\{0,\ldots,M-1\}$) is distributed as
$$
\Pr[\tilde{y} = z/M] = \frac{1}{M^2}\left(\frac{\sin(\pi M(z/M - x))}{\sin(\pi(z/M - x))}\right)^2 = \frac{1}{M}F_M(z/M - x).
$$
We used the properties of the sine function and the definition of $d(\cdot, \cdot)$ to get this equivalent form from Eq.~(\ref{eqerrorphasest}).
Hence $\tilde{y}$ is distributed as a discretized, renormalized Fej\'{e}r kernel re-centered at $x$.   (If $Mx \in \mathbb{Z}$, then $\Pr[\tilde{y} = x]=1$.) 

Recall that our proof required a sharper estimate than that given by plain phase estimation---we had to apply phase estimation three times.  Similarly, the convolution $g * F_n$ of $g$ with the Fej\'{e}r kernel fails to give a sufficiently close approximation to $g$ to prove Jackson's Theorem.  A sharper approximation is provided by the so-called \emph{Jackson kernel}, obtained by
squaring and renormalizing the  Fej\'{e}r kernel:
\begin{equation}\label{eqjacksonkernel}
J_n(t) = c F_n^2(t),
\end{equation}
where $c > 0$ is chosen so that $\int_{0}^{1}J_n(t)dt = 1$.  Jackson showed that $|(g * J_n)(x) - g(x)| = O(\omega_{1/n}(g))$ for all $x\in[0,1]$.  This $(g * J_n)$ is a trigonometric polynomial of degree $2(n - 1)$, which can be reduced to degree $\leq n$ by using $J_{\lfloor n/2 \rfloor}$ in place of $J_n$.  Note how squaring and renormalizing the Fej\'{e}r kernel has the effect of sharpening its concentration; this is somewhat analogous to the sharpening we achieve by taking medians-of-three in our quantum algorithms above.\footnote{We note that, if $\{p(z)\}_{0 \leq z \leq M - 1}$ is a distribution describing the outcome of a specific quantum measurement on a state $\ket{\phi}$, and if $q(z) = c p(z)^2$ where $c > 0$ is a normalizing factor, then the distribution $q(z)$ also has a somewhat natural interpretation in quantum terms.  Namely, suppose that we perform the measurement twice, on two identical (unentangled) copies of $\ket{\phi}$, yielding outcomes $z_1, z_2$, and condition on the event that $z_1 = z_2$.  Then $q(z)$ describes the probability that $z_1 = z_2 = z$ conditioned on $z_1=z_2$.  This can also be phrased in the language of \emph{quantum postselection}~\cite{aaronson:pp}, as we are postselecting upon the event $z_1=z_2$. The renormalizing constant $c$ in Eq.~(\ref{eqjacksonkernel}) is the reciprocal of the probability of this event.}

\section*{Acknowledgements}
\noindent
We thank two anonymous QIC reviewers for helpful comments.

%\bibliographystyle{unsrt}
%\bibliography{qc}

\end{document}